\documentclass[conference,letterpaper]{IEEEtran}
\IEEEoverridecommandlockouts
\usepackage[T1]{fontenc}
\usepackage[utf8]{inputenc}
\usepackage{times}
\usepackage{graphicx}
\usepackage{xurl}
\usepackage[hidelinks]{hyperref}
\ifdefined\pdfgentounicode
  \input{glyphtounicode}
\fi
\title{DruMR: A Feedback Driven Mixed Reality Air Drumming System%
\thanks{University of Minnesota College of Science and Engineering}}
\author{
\IEEEauthorblockN{1\textsuperscript{st} Viktor Radev}
\IEEEauthorblockA{\fontsize{11}{13.2}\selectfont University of Minnesota\\CSE\\Minnesota, United States\\Radev009@umn.edu}
\and
\IEEEauthorblockN{2\textsuperscript{nd} James Hebert}
\IEEEauthorblockA{\fontsize{11}{13.2}\selectfont University of Minnesota\\- Twin Cities\\CSE\\Minnesota, United States\\Heber212@umn.edu}
\and
\IEEEauthorblockN{3\textsuperscript{rd} Omar Elgohary}
\IEEEauthorblockA{\fontsize{11}{13.2}\selectfont University of Minnesota\\- Twin Cities\\CSE\\Minnesota, United States\\Elgoh003@umn.edu}
\and
\IEEEauthorblockN{4\textsuperscript{th} Akul Manoj\\Mundada}
\IEEEauthorblockA{\fontsize{11}{13.2}\selectfont University of Minnesota\\- Twin Cities\\CSE\\Minnesota, United States\\Munda057@umn.edu}
}
\begin{document}
\maketitle

\begin{abstract}
This paper introduces DruMR, a mixed reality (MR) drumming system designed to make learning drums more accessible, responsive, and feedback-driven. By combining Aerodrums hardware with the Meta Quest 3 headset, DruMR allows users to drum in a virtual environment without a physical drum set. Unlike existing VR drumming applications, DruMR emphasizes instructional value through quantitative feedback and future integration of AI-generated insights, refined quantitative and qualitative visualizations, reactive environment modes and more. Overall, this paper reviews the current state of the DruMR system, related works that influenced the systems development, planned user study comparing physical and virtual training, and future work to be added to the system.
\end{abstract}


\section{Introduction}
Learning to play an instrument comes with a few caveats: the cost of the instrument and lessons, the noise, and inability to play anywhere, to name a few. Drums have all of these issues, and thus can make these inaccessible to a beginner. To mitigate this, programs have been developed for VR that can allow you to play virtual drums. However, these programs lack instruction or feedback, making the barrier to learning drums still high. A way of delivering lessons should also be incorporated into the drumming programs so users will be able to improve their performance. The use of metrics, such as tempo consistency, and note accuracy, as well as tempo practice are a few ways to have lessons without another person present.

\medskip

This begs the question: can a drummer improve their playing without physical drums? This also leads to questions about what the most effective forms of feedback are for playing the drums. The purpose of this project is to answer these questions, and more specifically, find out how effective tailored metrics are for improving drumming skills.

\medskip

\section{Related Work}
Now we review related work in industry and the research space that uses VR/MR technologies that augment real-life musical instruments, help users try out instruments without needing the physical tools, and how feedback is vital in helping users retain interest in developing their musical endeavors.

\medskip

\begin{figure}[t]
\centering
\includegraphics[width=0.50\linewidth]{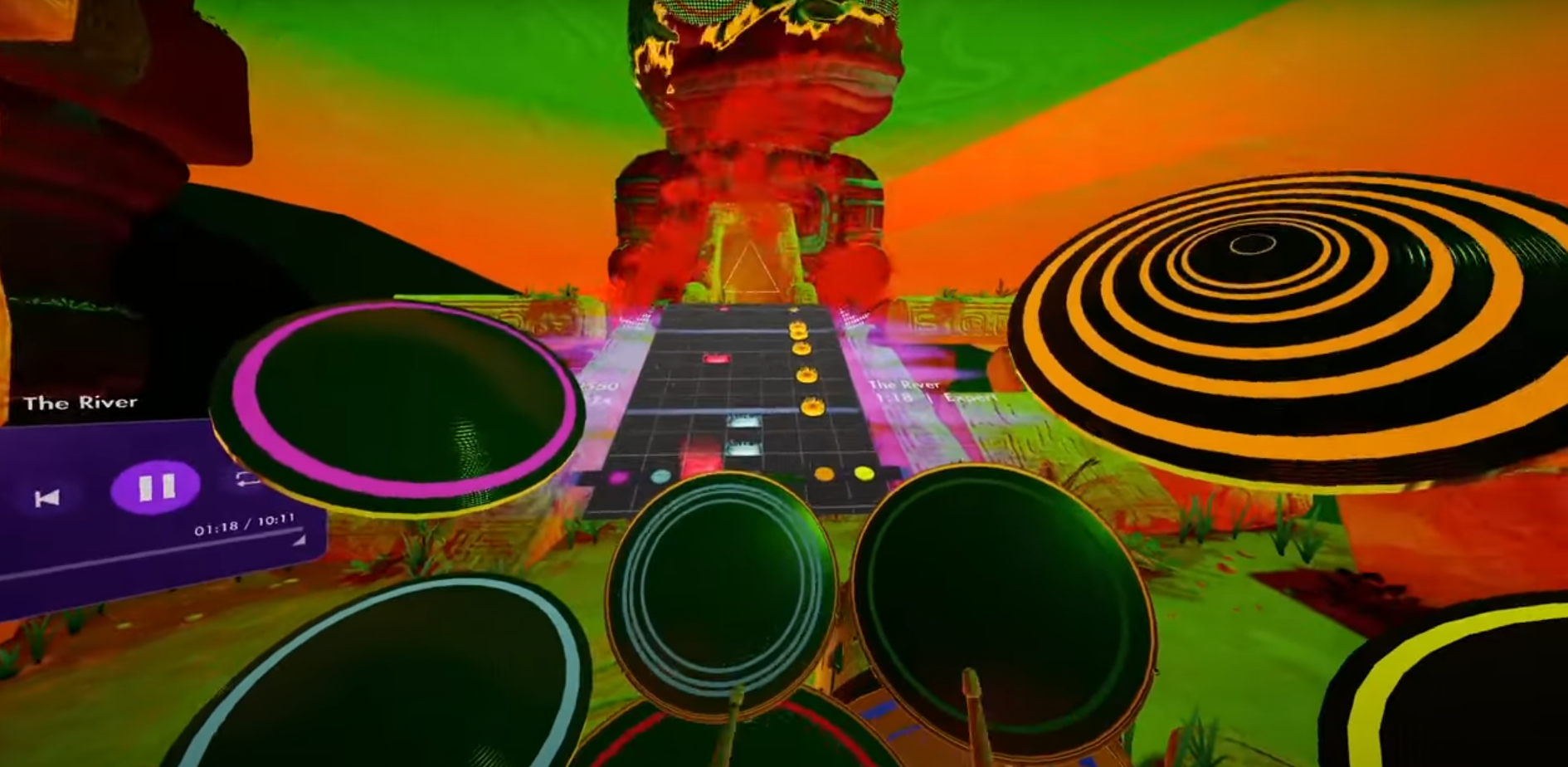}
\caption{This shows the various components paradiddle has implemented to make a MR/VR application to allow users to play drums without the need of a physical drum set. Instead of using a music sheet, it uses a guitar hero scrolling platform to indicate when it's users should hit a specific drum.}
\label{fig:1}
\end{figure}

\subsection{Paradiddle}
There are plenty of VR/MR applications that allow users to experience what it is like to play various instruments without the need for a physical instrument. However, Paradiddle is one of the more recognizable VR/MR applications for allowing users to augment their physical drum set or to play drums without the need for a physical drum kit.

\medskip

The main attraction of this application is that it allows users to practice drumming with just a headset and controllers. This application provides a drumming experience to its users by having them pick from a selection of preset songs. Then this loads in a drum kit into their environment, which can be hit with drumsticks tethered to the user's controllers. Once the users position their drum kit into their 3d space as needed, the song they chose to play will be rendered onto a guitar hero style visualizer, which tells the users when they should hit a particular drum.

\medskip

In addition to allowing users to build muscle memory and hear what it would be like to play physical drums without a physical drum set, it also allows users to take in input from their electric drum kits and then add interesting visuals to their drum kits when they are hit. The user places ghost drums into the drum kits, and when they are struck, a small visual effect emanates from the ghost drum, and in the end, it will track the user's score on the predefined tracks provided in this mode. One limitation of this mode is that the user must use the mapping provided in the pre-downloaded library of Paradiddle. However, an external software called ''paredit'' was created to allow users to upload custom songs to the Paradiddle application (2:10 - 2:22) \cite{ref1}.

\medskip

The meta quest review page for Paradiddle reveals some things that could be improved in future systems that allow users to play musical instruments in a VR/MR reality setting. GoldMaca7842 on the Paradiddle platform stated, ''Amazing game, super fun. I would suggest only buying this if you are a drummer, but it's also great for beginners. There are 6 lessons or 10, I can't remember, but if you do the drum lessons, it leaves you kind of on a cliffhanger. It doesn't teach you much, but I love this game. . . '' \cite{ref6}

\medskip

Another review on the system provided by an amateur drummer on YouTube stated, ``Still a little bit gimmicky, definitely not something like Beat Saber responsive, where everything is just very responsive.'' \cite{ref2}

\medskip

Reading more than fifty reviews of a similar nature shaped the research motivations mentioned in the introduction. In addition to making a system tailored to helping provide effective feedback to its users, it also pushed the system to use specialized hardware to provide a more responsive drumming experience to its users, and that's the motivation behind why we are using the AeroDrum System.

\medskip

\subsection{Feedback Strategies for Music Education}
A discussion on one of the most effective ways to motivate musicians toward maintained engagement and growth in learning music is to provide them with tools and an environment that facilitates a flow state experience. A flow state arises when users are provided with immediate feedback, deep concentration, and control over their situation. Music Games in Education: \cite{ref3}

\medskip

Looking into a system called Mimi (a multi-modal interactive musical improvisation system) it takes in input from its user and ''displays a scrolling piano roll notation, showing past, present, and future notes, with a central bar marking the present.'' It also includes potential notes the musician should consider by implementing a ''factor oracle'' that adds additional notes into the future section of the tape for the musician to consider. The system constantly provides visual feedback, and based on what the user wants, they may accept the oracle's suggestion or continue playing their music as initially planned. \cite{ref4}

\medskip

The system attempts to provide an environment where musicians experience flow by being open-ended and constantly suggestive, thus supporting improvisation without overwhelming the user. Providing such an experience is vital for sustaining motivation and has been implemented in other creative musical systems, including Mimi. This design is part of the motivation for implementing a Large Language Model (LLM) in the future to provide realtime, personalized feedback based on metrics collected throughout a session, enabling performers to receive guidance on what to improve upon after each performance. In addition to implementing an LLM into the system to summarize the users performance, realtime feedback can be provided with a virtual reality experience featuring a reactive audience, further reinforcing the sense of immersion and live feedback.

\medskip

In addition to making a system reactive to create a flow state environment for users, feedback must also be well structured and provide quantitative feedback for the users to evaluate what they may need to improve upon. Moreover McPherson et al. \cite{ref5} explains, ''feedback that is given at or just above the level the student is working at . . . should also aim to move them upward through the various levels required to achieve mastery'', thus the metric system that is going to be developed for this system will go beyond just giving a score, or providing them with a simple hit/miss counter Feedback in Music Performance Teaching.

\medskip

Instead, it will include structured metrics such as true positives, false positives, a heatmap of where drums are striked, average latency of note hits, and a visual timeline highlighting when mistakes occurred. Additionally, adding color-coded drums based on error that occur while the user plays, will simplify beginners' realtime feedback while continuing to offer the in-depth analysis spoken about before. These visualizations are only made possible with the affordances of Mixed Reality, such as allowing for custom visual overlays development and creating a responsive system that reacts to in-game events.

\medskip

Therefore, the system to be developed will be specifically tailored to delivering deep yet accessible analysis of user performance. By integrating quantitative data with meaningful visual feedback and reactive guidance, it aims to motivate users across all skill levels to keep improving. Most importantly, it fosters an environment that helps users enter a flow state and making practice an intrinsically rewarding experience. These benefits go beyond the original motivation of the creation of products such as the aerodrum:

\medskip

''My personal motivation for developing this, when I was young I used to play drums and had problems with the neighbors, so I had an arrangement with the neighbors where I could only play 2 hours a day and had to give up when I came home from college and work in the evening and it was too late to play drums.'' \cite{ref7}

\medskip

\begin{figure*}[t]
\centering
\includegraphics[width=1.00\linewidth]{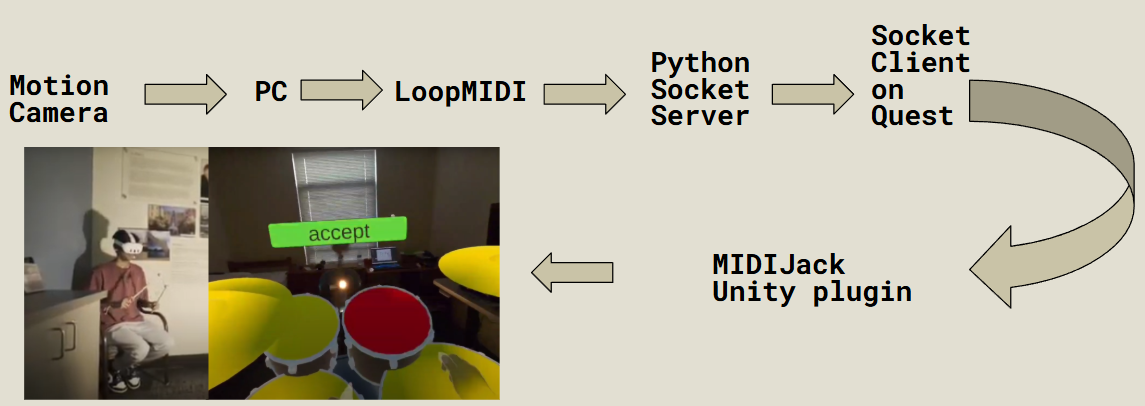}
\caption{Figure 1: This figure shows the standard pipeline used for this project. First a motion camera captures the position of the drumsticks of the users. Then the AeroDrum software converts this position data into a midi note, which is then taken by Loop MIDI and transferred to a UDP socket server running on the same device as the Aerodrum software. The socket server then takes those MIDI notes and then sends them to the users headset via a network connection, and the meta quest then lights up the corresponding drum red.}
\label{fig:2}
\end{figure*}

\section{Methodology}
\subsection{Hardware}
In order to have a system that is both more accessible than a drum set, and capable of precise tracking, the system uses the Meta Quest 3 and the AeroDrum. With the VR space, users no longer need to worry about having a physical drum kit with them.

\medskip

The Meta Quest 3 is a standalone VR headset that supports tracking, high-resolution rendering, and wireless operation. It is the primary display and interaction platform for our application. The headset runs the Unity application natively, allowing users to see and interact with a virtual drum kit in 3D space without adding additional external hardware onto the headset to provide a more comfortable setup.

\medskip

The AeroDrum system is a camera-based drumming interface that allows users to play a virtual drum kit using physical drumsticks with reflective markers. Instead of having users buy a physical drum set it instead uses motion tracking techniques via a high-speed camera (e.g., PlayStation Eye) and infrared illumination to detect drumstick position and velocity.

\medskip

Each drumstick has a reflective tip, and a small set of foot markers (e.g., for bass drum) can be placed on the user's shoes. The camera, combined with computer vision software, tracks the 3D position and velocity of each stick, and maps the motion into virtual drum hits. AeroDrums can produce real-time audio feedback through its own sound engine or output standard MIDI messages.

\medskip

One caveat of the Aerodrum hardware is that the users must play in a dimly lit room with minimally reflective objects present in their background. There is a light attached to the camera which allows for aerodrum software to track the reflective markers and any light not coming directly from the camera interferes with the reflection based tracking of the system.

\medskip

\subsection{Software}
\subsubsection{Real-Time MIDI Capture and Transmission}
Using both a UDP socket server, and the Mido library, the MIDI output generated from the AeroDrums is streamed to the Meta Quest. Three pieces of information are streamed every time the drums are hit: note, velocity, and the timestamp as seen in Figure 2. These are later used for metrics. In order to ensure there are no double triggers, a queue allowing for no more than 33 hits per second filters those out.

\begin{figure}[!ht]
\centering
\includegraphics[width=1.00\linewidth]{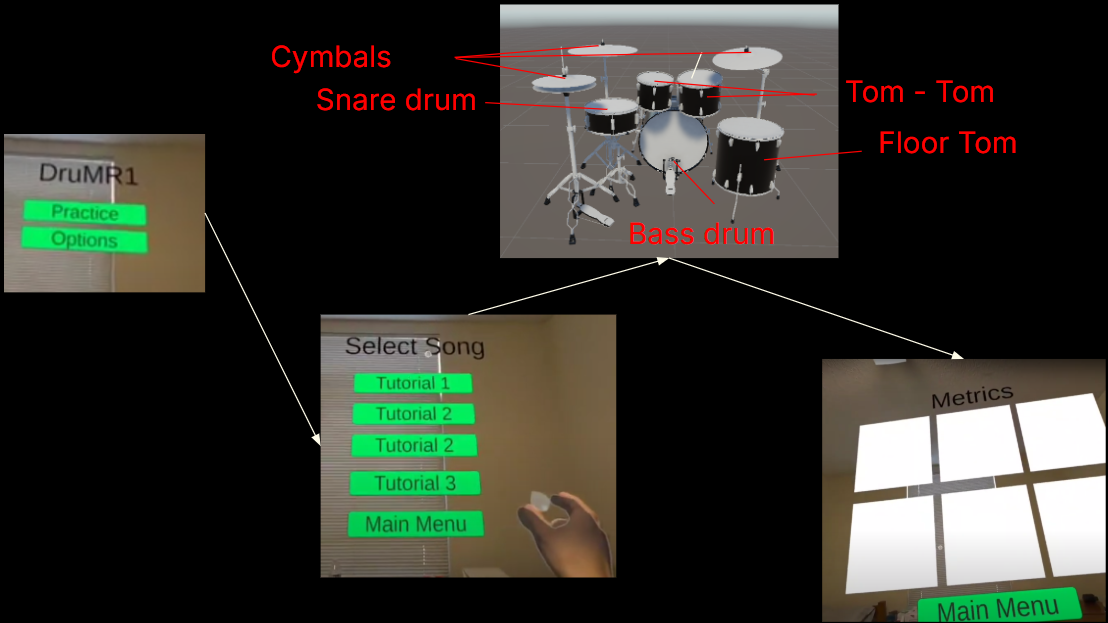}
\caption{Visual description showing the optimal game play loop for the system. First the user chooses whether they want to practice, (work on tempo is a work in progress mode), or change some visual options. The the user selects from a set of preset songs. This then loads up the 3D drum set into their view which can be repositioned in their environment as needed. Finally they can accept their drums positioning which then starts a countdown. Finally, a metrics panel pops up with various pieces of data.}
\label{fig:3}
\end{figure}

\subsubsection{Unity Gameplay Loop and Input Handling}
The gameplay loop begins with the user in the main menu. From there, the user can select tempo Practice, free play, and drill practice. Selecting one will allow the player to position the drums in 3D space according to where they are created by the AeroDrums. When the user is satisfied with the calibration, the can start playing, and the drums will light up red when struck. Once the session is over, the metrics are displayed to the user.

\subsubsection{Metrics Evaluation}
\begin{figure}[!ht]
\centering
\includegraphics[width=0.75\linewidth]{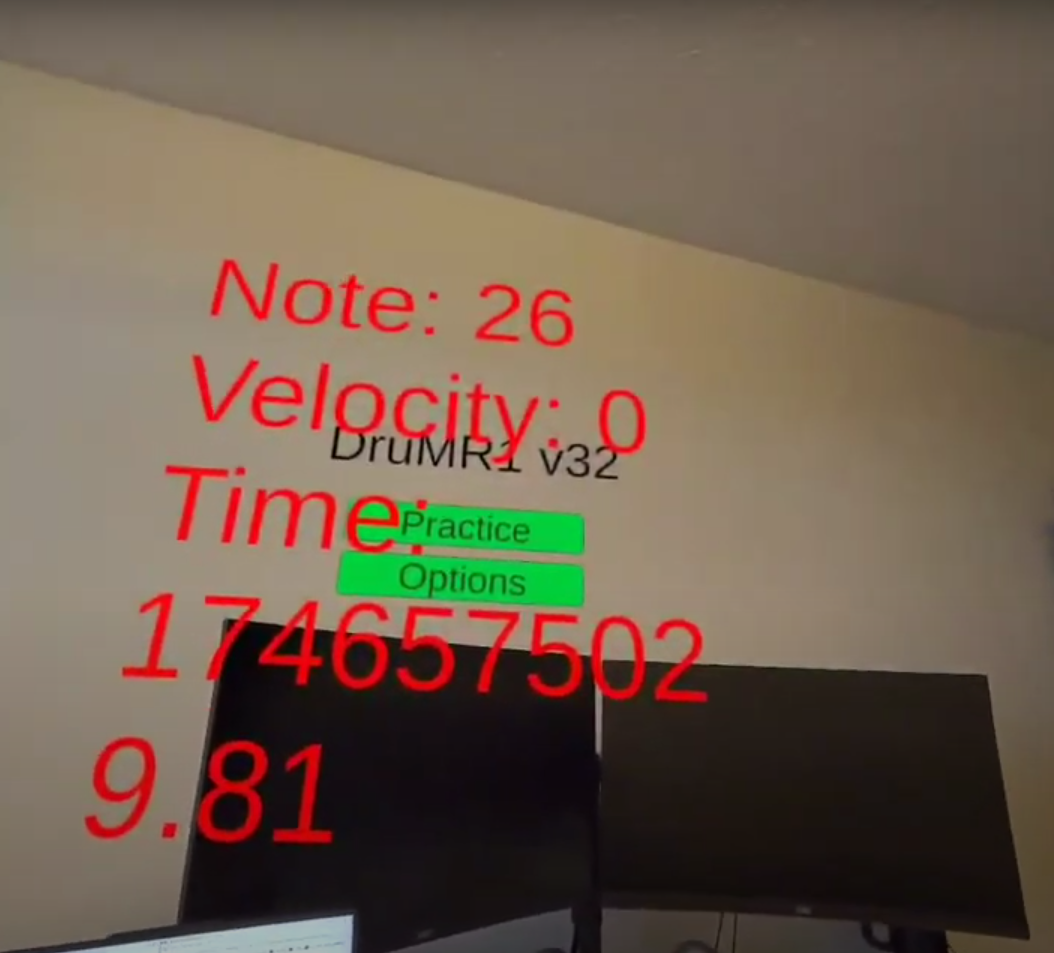}
\caption{Image showing Note, Velocity, and Time being extracted from a MIDI note being sent from the AeroDrum software to the Meta Quest via a network connection. Extracting note data in runtime will allow for metrics to be developed such as a timeline providing users with info on what part of a song they should continue practicing, real time hit/miss indicators, and more.}
\label{fig:4}
\end{figure}

The metrics are calculated by comparing the note data to a predefined MIDI file. The data displays the note accuracy - the overall amount of correct hits, and the tempo accuracy - the overall consistency of staying on beat. Since there is some latency, a tolerance of $\pm$150 ms is used. These are displayed after each session.

\medskip

\section{User Evaluation}
The plan is to do a user study with two groups. One group will start by practicing on the drums, and the other group will begin with DruMR. Each group will be given 10 minutes time to familiarize themselves with the equipment and be trained by a professional drummer, and then be tasked with performing a drum beat as a test for 10 minutes. After a set period of time, each group will demonstrate the beat they learned and play for 5 minutes after being trained on the real and DruMR system, which will then be judged by the professional drummer. In addition to the professional drummers judgement, midi sounds from the real drums will also be compared to the song directly with the use of various recording tools. The professional will not be told who started with DruMR and who started on physical drums. Each person will secretly receive a rating, and that will be used to compare the two groups.

\medskip

After the users complete the series of tests and training, they will have to complete a few surveys and a post p- study evaluation. The first survey they must complete will be the NASA TLX survey which measures perceived workload (mental, physical, effort, frustration, etc.). The second survey to be completed will be the system usability scale which rates ease of use and usability of the DruMR system. The third survey will be the Likert Scale for the DruMR's usability. Finally, the Post-Study Evaluation will have an open-ended interview with questions such as, Did the VR training help, how did the DruMR system compare to the real drum training, What could be improved on the DruMR system? Finally, an objective performance metric will be conducted on the users which will have a quantitative analysis of improvement post-training, comparing pre- and post-training scores via the midi files that are captured during their play sessions.

\medskip

\section{Future Work}
To make DruMR a better platform for learning , there are plans to implement several features. First is simulating a live audience, which will react to the drummer's performance. This will help users with performing on stage. Feedback is another method that will be implemented. Based on how a user plays, a LLM will tailor feedback designed to help them learn. The metrics will also be improved to locate which portion of a song the user messed up the most. From there, they can practice that portion until satisfied with their results. The last feature to implement will be a heatmap, which will overlay on the drums after the session. The heatmap will show where the drums were hit the most, as well as how hard by using a color gradient.

\end{document}